\documentstyle[11pt,newpasps220,twoside,epsf]{article}
\def\edcomment#1{\iffalse\marginpar{\raggedright\sl#1\/}\else\relax\fi}
\reversemarginpar

\begin{document}
\title{A 3D Optical Spectroscopy Study of Low Surface Brightness Galaxies}
 \author{L. Chemin$^{1,2}$, P. Amram$^3$, C. Carignan$^1$, C. Balkowski$^2$, W. van Driel$^2$, V. Cayatte$^2$, O. Hernandez$^{1,3}$, 
J. Boulesteix$^3$,  M. Marcelin$^3$}
 \small
\affil{$^1$ Dept. de Physique, Universit\'e de Montr\'eal, C.P. 6128, Succ. Centre-ville, Montr\'eal, Qc, Canada, H3C 3J7}
\affil{$^2$ GEPI, Observatoire de Paris, 5 Pl. J. Janssen, 92195 Meudon, France}
\affil{$^3$ Observatoire Astronomique de Marseille-Provence, 2 Pl. Le Verrier, 13248 Marseille, France}
\normalsize
%\vspace*{-0.35cm}
\begin{abstract} 
We present H$\alpha$ emission line velocity fields of two Low Surface Brightness galaxies  (\small LSBs\normalsize)
- UGC 628 and UGC 5005 - obtained using Fabry-Perot  (\small FP\normalsize) interferometry observations 
at the Canada-France-Hawaii telescope.  Our goal is to study the dynamics of \small LSBs\normalsize.
\end{abstract}
\vspace*{-1.0cm}
\section{Introduction}
\vspace*{-0.25cm}
Low Surface Brightness galaxies are ideal laboratories 
to study the dark matter (\small DM\normalsize) properties and it is still debated 
whether their dark haloes are cuspy or core-dominated 
(de Blok et al. 2001, Marchesini et al. 2002, Swaters et al. 2003). 

\small FP\normalsize\ interferometry is a powerful tool for studying the  \small DM\normalsize\
properties of  \small LSBs\normalsize\ for two major reasons:\\
\hspace*{1.5cm} $\bullet$ the accurate determination of the \small DM\normalsize\ distribution in galaxies 
requires high angular resolution observations (see eg. Blais-Ouellette et al. 1999),\\
\hspace*{1.5cm} $\bullet$ 3D spectroscopy like \small FP\normalsize\ 
provides more accurate kinematical parameters and rotation curves than long-slit observations 
as it samples the whole velocity field of a galaxy. 
 
Using a new generation of detectors with unprecedented sensitivity (Gach et al. 2002), we began a campaign of 
 \small FP\normalsize\ observations of \small LSBs\normalsize\ in order to study 
 the kinematics of the ionized gas and their dynamics (Chemin et al. in prep). 
 
\vspace*{-0.5cm}
 \section{Observations, results and discussion}
 \vspace*{-0.25cm}
The H$\alpha$ emission line in UGC 628 and UGC 5005 was observed on October 2002 and April 2003 (resp.) at the Canada-France-Hawaii telescope, 
using \small FaNTOmM\normalsize\footnote{http://www.astro.umontreal.ca/fantomm}, a scanning  \small FP\normalsize\ interferometer coupled 
with a photon counting camera. 
The angular resolution is 0.49" and the spectral sampling is 7 km/s for UGC 5005 and 16 km/s for UGC 628. Exposure times 
are of the order of 3h per galaxy.  

Figures 1 and 2 present some provisional results. 
Figure 1 shows the velocity fields of UGC 628 and UGC 5005. Figure 2 
shows the  mass models (Blais-Ouellette et al. 2001) for UGC 5005  with the pseudo-isothermal (\small ISO\normalsize)
sphere (left) and the \small NFW \normalsize\ (right) \small DM\normalsize\ halo profiles. 
  
A dozen objects were observed as of date. 
The high angular resolution \small FP\normalsize\ observations will allow a detailed study of 
\small LSBs\normalsize\ velocity fields and rotation curves and their \small DM\normalsize\ content. 
We will also study the effects of non-circular motions on their mass distribution (Chemin et al. in prep).

\begin{figure}
\plottwo{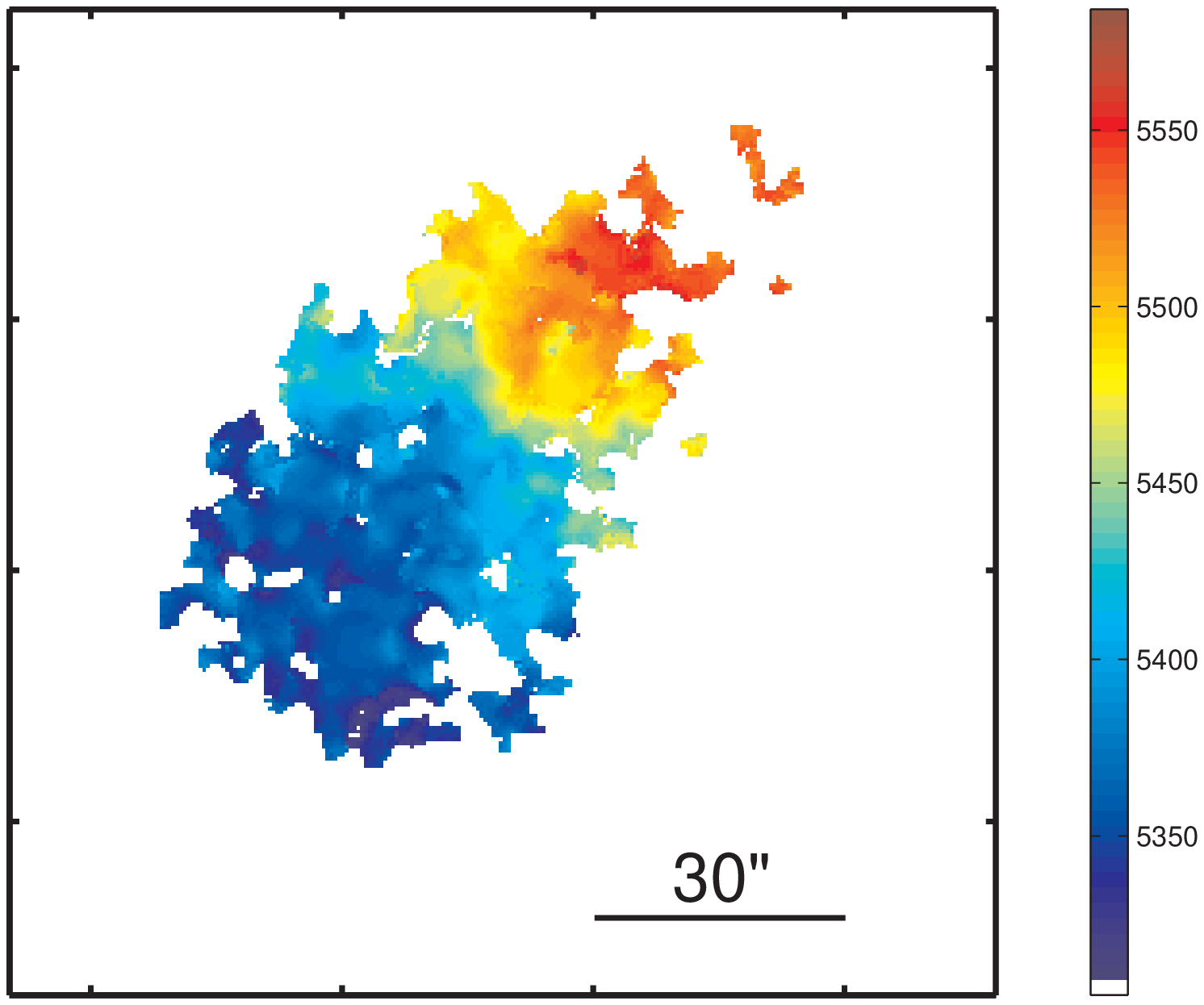}{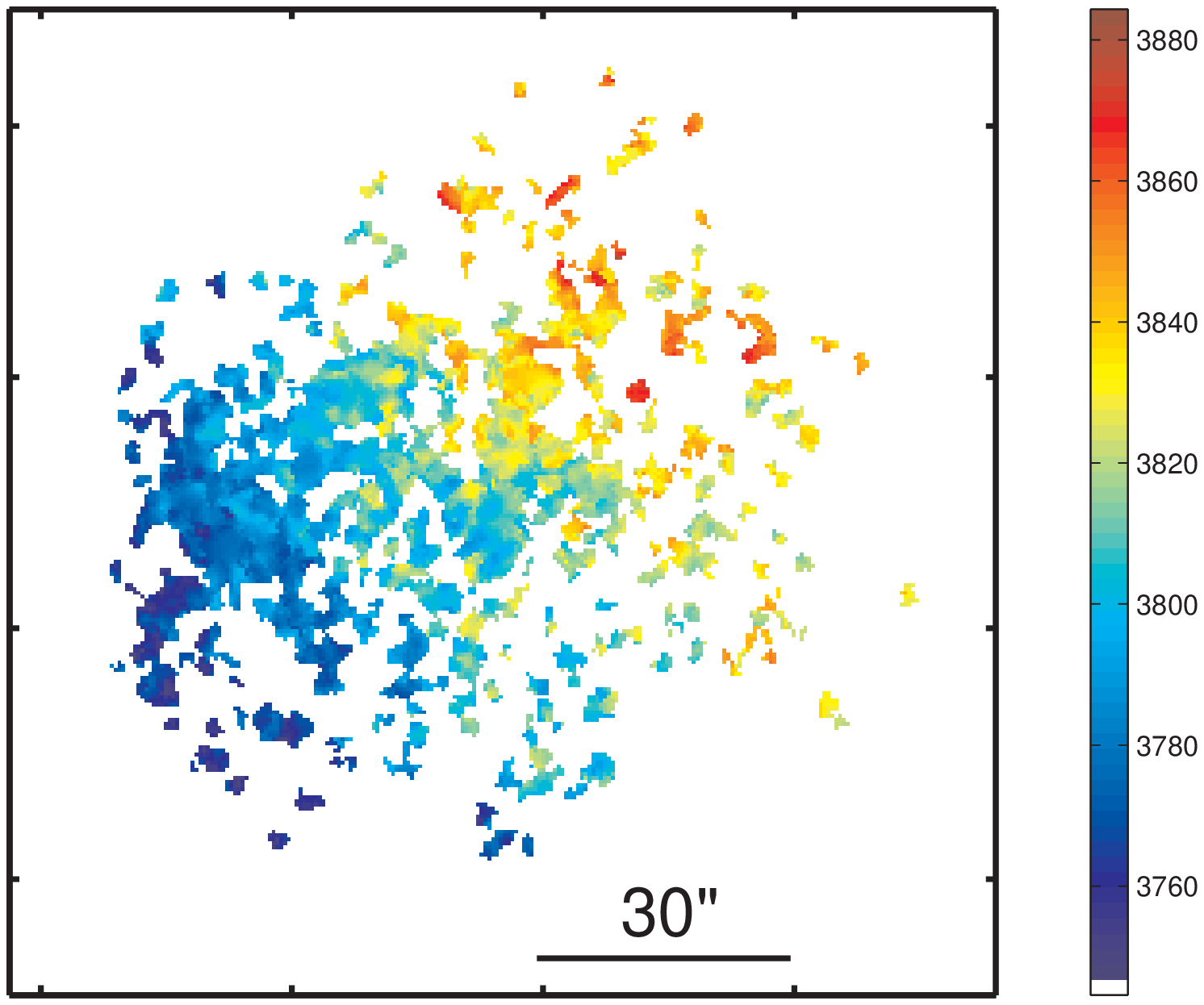}
\caption{\small H$\alpha$ velocity field of UGC 628 (left) and UGC 5005 (right). Heliocentric velocities are in km/s.
\normalsize}
%\end{figure}
%\begin{figure}
\plottwo{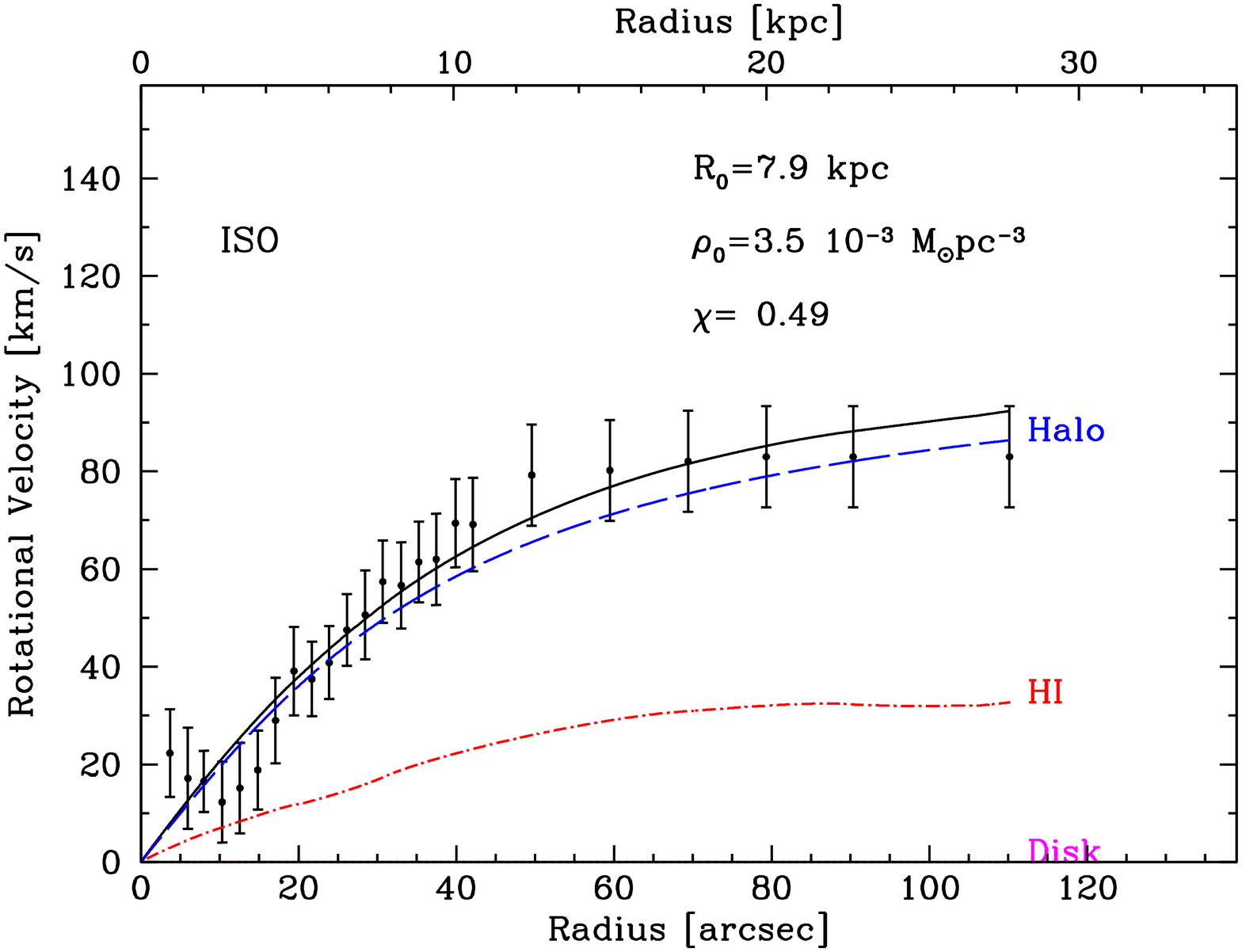}{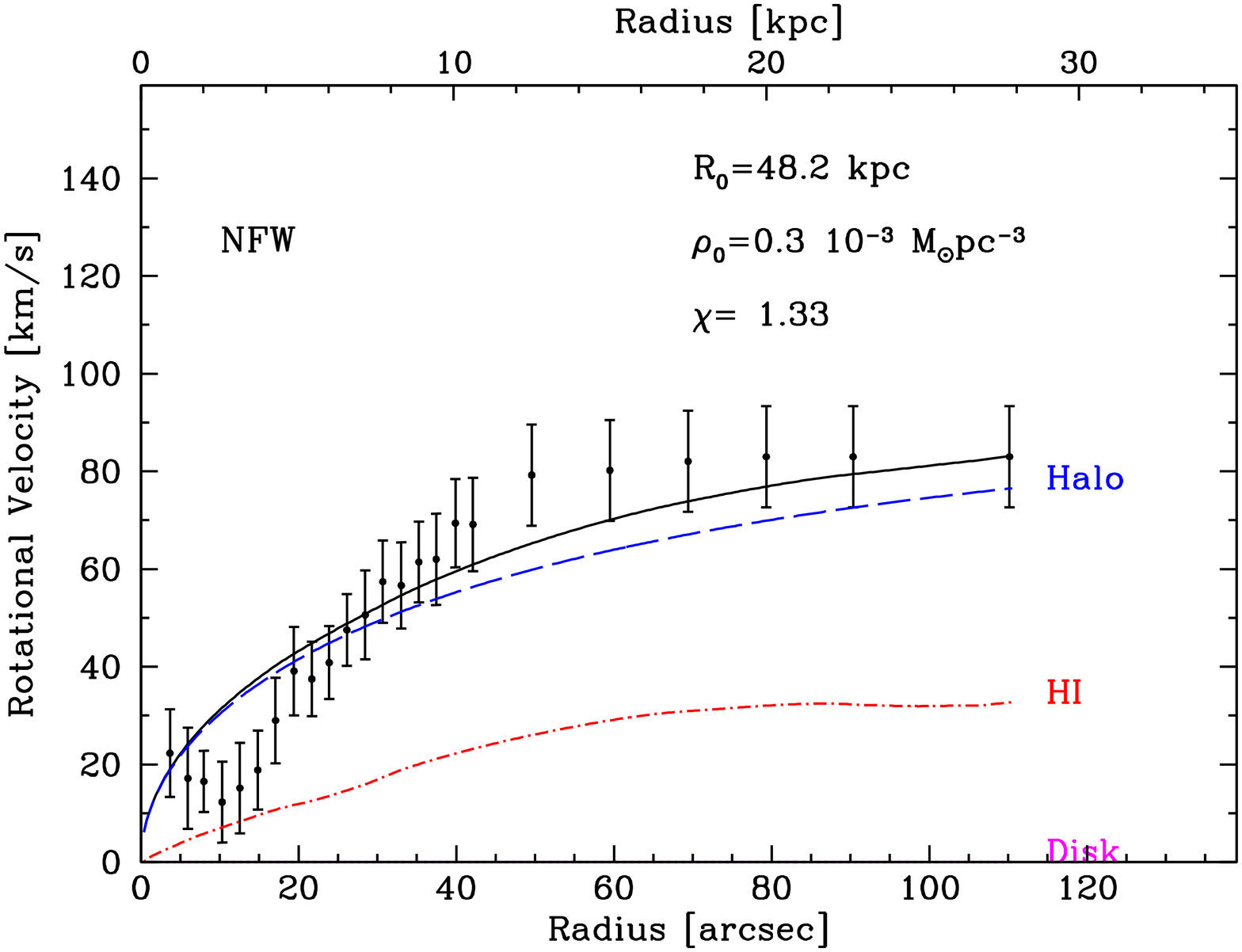} 
\caption{\small  Best fit mass models of UGC 5005. Full circles represent the observed rotation curve, combining  
our H$\alpha$ points for $R < 45$\arcsec, and HI observations elsewhere (van der Hulst et al. 1993).  A 
 blue dashed line and a red dotted-dashed line are the rotation curves of the halo and the gas components (resp.). 
 A full line represents the total fitted rotation curve.
 For each model, the deduced halo parameters and the $\chi$ value are indicated. \normalsize \vspace*{-0.2cm}}
\end{figure}

\end{document}